\documentclass[twocolumn,showpacs,preprintnumbers,amsmath,amssymb]{revtex4}
\usepackage[english]{babel}

\usepackage{graphicx}

\newcommand{\beq}{\begin{equation}}
\newcommand{\eeq}{\end{equation}}
\newcommand{\bma}{\begin{math}}
\newcommand{\ema}{\end{math}}
\newcommand{\beqa}{\begin{eqnarray}}
\newcommand{\eeqa}{\end{eqnarray}}

\def\opone{\le\textbf{}\textbf{}avevmode\hbox{\small1\kern-3.8pt\normalsize1}}

\begin{document}

\title {Microscopic theory of the quantum Hall hierarchy}

\author{E.J. Bergholtz}
\author{T.H. Hansson}
\author{M. Hermanns}
\author{A. Karlhede}

\affiliation{Department of Physics, Stockholm University,\\ AlbaNova University Center, SE-106 91 Stockholm,
Sweden\\
}

\date{\today}

\begin{abstract}

We solve the quantum Hall problem exactly in a limit and show that the ground states 
can be organized in a fractal
pattern consistent with the Haldane-Halperin hierarchy, and
with the global phase diagram.
We present wave functions for a large family of 
states, including those of Laughlin and Jain and also for states recently observed by Pan {\it et. al.}, 
and show that 
they coincide with the exact ones in the solvable limit. We submit that they establish an
adiabatic  continuation of our exact results to
the experimentally accessible regime, thus providing a unified approach
to the hierarchy states.

\end{abstract}


\maketitle

A key concept in quantum Hall (QH) physics is that of an incompressible electron liquid. 
In the integer effect, the formation of this liquid
can be understood in terms of independent electrons moving in a magnetic field in the 
presence of a small but crucial amount of disorder\cite{laughlin81}. 
In the experimentally very 
similar fractional effect, the electron liquid is formed through the electron-electron interaction. 
The fractional liquids, the simplest of which are well understood in terms of Laughlin's wave 
functions\cite{laughlin83}, are highly correlated quantum systems with 
remarkable properties.

\begin{figure}[h!]
\begin{center}
\resizebox{!}{65mm}{\includegraphics{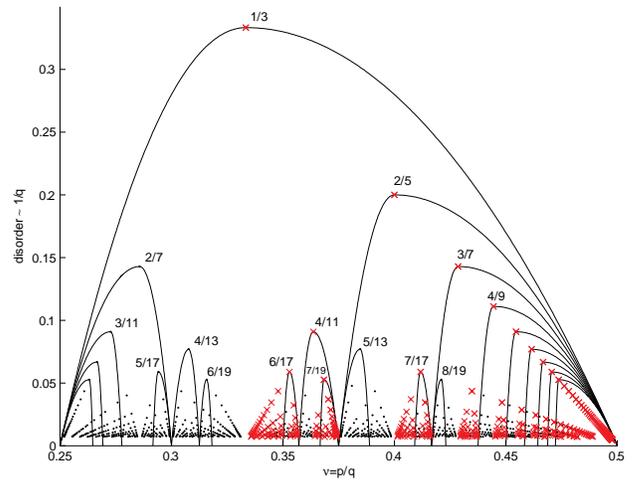}}
\end{center}
\caption{\textit{{\small  {\bf Global phase diagram.}
The global phase diagram in the filling factor-disorder  plane  for $1/4\le \nu \le 1/2$ (other regions are similar). Crosses mark  fractions 
where wave  functions are constructed using conformal field theory.
}
}}\label{figure4}
\end{figure}

There are two alternative microscopic approaches to the fractional liquids\footnote{
We consider here only spin-polarized abelian QH states.}---the hierarchy 
and composite fermions. In the former, successive condensation of the fractionally charged quasiparticles 
leads to a hierarchy of ever more complex QH-states\cite{haldane83,halperin84}. 
In the latter,
the electrons bind magnetic flux quanta to form new particles, 
composite fermions, that see a reduced magnetic field, and the fractional effect is an integer effect of these new particles\cite{jain89,jainbook}. 
However, neither composite fermions nor the hierarchy has a solid theoretical foundation, and they 
may well be complementary views of the same phenomena rather than mutually excluding 
descriptions\cite{read90,greiter}.  The hierarchy construction is closely related to the global phase diagram, see  Fig. \ref{figure4}, which provides an overall picture of the QH-states.   This phase diagram, which can be derived using a Chern-Simons-Ginzburg-Landau approach\cite{global}, and exhibits an intriguing modular symmetry\cite{lutken}, 
is supported by experiments, 
but deviations have also been reported\cite{jiang}, and its precise status remains an important open problem.

A great advantage of the composite fermion scheme is that it provides very good and explicit wave functions for the Jain states with 
$\nu=n/(2kn\pm 1)$, $n,k=1,2,\dots$, which can be interpreted as integer QH-states of composite fermions. 
Until recently, all observed states 
were of this type; in 2003, however, experiments on ultra clean samples revealed 
states at other fractions, such as $4/11$ and $6/17$\cite{pan}.   These new states point towards a fractal, self-similar, structure of 
states in the lowest Landau level\cite{mani}. 
At present, there are no agreed upon wave functions  for these states and the proper interpretation 
of them is under debate. 
In this Letter, we give theoretical support for the global phase diagram and present an explicit realization of the hierarchy in a well-defined and solvable limit. We also report explicit and testable 
wave functions for the states that are obtained by successive condensation of quasielectrons---these wave functions agree with 
our exact solution in the solvable limit, and naturally encompass both the
Laughlin/Jain  wave functions and states such as 4/11, reported in Ref. \onlinecite{pan}. 
For other attempts to construct hierarchy wave functions, some of them with explicit reference to composite fermions, see {\it e.g.} Refs. \onlinecite{greiter,other}.

\begin{figure}[h!]
\begin{center}
\resizebox{!}{65mm}{\includegraphics{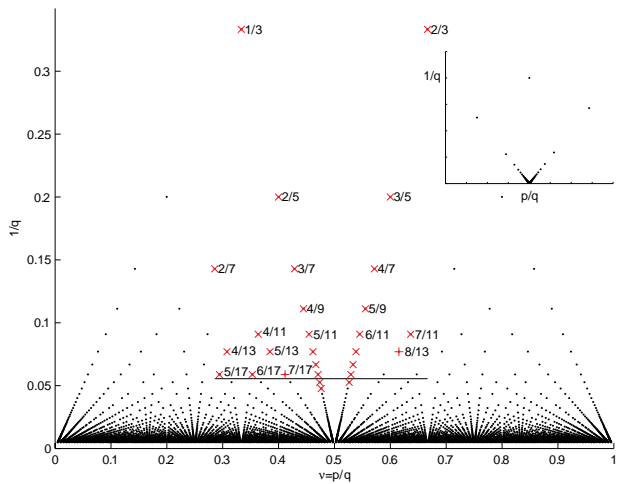}}
\end{center}
\caption{\textit{{\small  {\bf Observed states and fractal structure.}
For each rational filling factor, $\nu=p/q\le 1, \, q$ odd, 
there is a unique hierarchy state and its stability increases monotonously with $1/q$. 
The fractal structure of states is manifest. 
States in the region $2/7\le \nu  \le 2/3$ reported in Ref. \onlinecite{pan} are marked by crosses; plusses mark fractions were we infer a possible 
weak signal from the data in Ref. \onlinecite{pan}. The horizontal line marks the extent in $\nu$ 
of the experiment and is 
a line of constant gap and the approximate boundary for the observed states. 
The inset shows the  structure of hierarchy states: At each $\nu=p/q$, $q$ odd, there is a state with gap $\sim 1/q$ and 
quasiparticles with charge $\pm e/q$. When these condense two sequences of states approaching $p/q$ with decreasing gap are obtained. 
}
}}\label{figure3}
\end{figure}

There is  a simple and striking consequence of the experiment in Ref. \onlinecite{pan}: QH-states 
are observed at all filling factors $\nu=p/q$ for $q\le q_0\approx   17$ in the 
experimental range of $\nu$ (here $q$ is odd and $p,q$ are relatively prime), see Fig. \ref{figure3}. 
A QH experiment is performed on a sample with a certain 
amount of disorder---the lower the disorder is, the more fractions are observed. This defines a notion of stability for a state: 
a more stable state is seen at higher disorder than a less stable state. Our interpretation of Ref. \onlinecite{pan} is that: 1) there is a 
QH-state at any $\nu=p/q \le 1$, $q$ odd, and 2) the stability of the state decreases monotonously with increasing $q$
 in agreement with the hierarchy prediction\cite{halperin84}.

The statement that there is a QH-state for each $\nu=p/q\le1$ needs to be qualified. 
We introduce a  parameter $L$ such that the experimental regime is obtained as $L\rightarrow \infty$. When $L\rightarrow 0$ the 
problem is exactly solvable and the ground state and its quasiparticle excitations have 
all the qualitative properties of a QH-state. If there is no phase transition as $L$ increases, then the QH-state is observed 
at this filling factor. However, phase transitions may occur leading
to other states such as  Wigner crystals where the repulsion freezes the electrons in a regular lattice,  
striped states where the electron density varies periodically or a Fermi gas (for even denominators $q$).

To give a detailed argument for the propositions in the introduction, we  proceed in three steps. 
First we solve the interacting spin-polarized many-electron system exactly 
for each filling factor $\nu=p/q\le 1$  in a certain mathematically well-defined limit. We obtain the ground state as well as 
the fractionally charged quasiparticle excitations. Furthermore, in this limit
the hierarchy construction of the QH-states is manifest, and 
the stability of the states decreases monotonously with $q$. 
Second, 
for each fraction where the state is obtained by (successive) condensation of 
quasi{\em electrons} (as opposed to quasi{\em holes}), see Fig. \ref{figure4}, we present explicit wave functions 
for the ground state and wave functions for the quasiparticle excitations can also be obtained. 
These wave functions---which are in the lowest Landau level---are obtained by a unique 
and natural construction that exploits the intriguing relation between the QH effect and 
conformal field theory (CFT). They are in 
one-one correspondence with the exact solutions, and reduce to these 
 in the solvable limit. Furthermore, the fractional charge
and fractional statistics of the quasiparticles are reflected in the algebraic properties of 
the CFT-operators by which they are created. 
Third, we argue that this construction 
establishes the adiabatic continuation of the results in the solvable limit to the experimentally accessible regime (at fractions where a QH-state is observed). 

The solvable limit is obtained by considering the two-dimensional electron gas on a cylinder with circumference $L$, 
and choosing the one-electron states centered along rings around the cylinder---these states are gaussians with width of order one
magnetic length $\ell =\sqrt{ \hbar c/eB}$ along the cylinder.
This maps the QH problem onto a one-dimensional lattice problem with lattice constant
$2\pi \ell^2/L$. 
A basis of many-electron states is given by $|n_1n_2,\dots \rangle$, where $n_k=0,1$ depending on whether 
site $k$ is empty or occupied. 
When $L/\ell \rightarrow 0$ the overlap between different one-electron states vanishes and 
the energy eigenstates are 
simply the  states  where the electrons occupy fixed positions, $|n_1n_2,\dots \rangle$, and the
ground state is the one that minimizes the electrostatic repulsion\cite{bk2}.
At $\nu=1/3$, this is obviously the state where every third site is occupied\cite{tt}. For general filling factor, 
$\nu=p/q$, the ground state is a gapped crystal---or Tao-Thouless (TT) state---with $p$ electrons in a unit cell of length $q$. 
For example, at $\nu=1/3, 2/5, 3/7,\dots$ the unit cells are $100, 10010\equiv 10_210,1001010\equiv 10_2(10)_2\dots$. 

The lowest energy charged excitations at $\nu=p/q$ are quasielectrons and quasiholes with charge $\mp e/q$, where $-e$ is the electron 
charge; in the thin limit these are 
domain walls in the TT-state and the charge is determined by the Su-Schrieffer counting argument\cite{su,bk2}. For example, 
the excitations at $\nu=1/3$ are obtained by inserting or removing 10.
When the filling factor is gradually increased away from 1/3 a gas of
such quasielectrons of increasing density is formed. These repel each other and condense to form new ground states. Eventually,  one 
excitation per unit cell has been added and the new ground state has unit cell 10010---the filling factor is then 2/5.
However, before this happens lower density condensates, with one quasielectron per $2k-1$ unit cells 100 will form, giving new 
ground states with unit cells $(100)_{2k-1}10$  at filling 
factors $\nu=2k/(6k-1)$,  $k=1,2,\dots$. 
Similarly, decreasing the filling factor away from 1/3 gives the 
ground states with unit cells $(100)_{2k-1}1000$ at $\nu=2k/(6k+1)$. This pattern is general: The TT-state at 
$\nu=p/q$ is the parent state for two sequences of daughter states that approach $\nu=p/q$ from above 
and below and  are obtained by condensation of decreasing densities of quasielectrons and quasiholes respectively, see the inset in Fig. \ref{figure3}. 
For details and proofs  we refer to Refs. \onlinecite{bk2,bk3}.

To summarize, at each $\nu=p/q\le 1$, $q$ odd, there is a TT-state, which we argue below
develops into a QH-state as $L\rightarrow \infty$, and these states are formed from other 
TT-states by condensation of quasiparticles. 
This may also be interpreted as the quasiparticles filling an effective Landau level as one goes {\it e.g.}  from $\nu=1/3$ to $\nu=2/5$. 

The QH effect is destroyed by disorder---at a certain amount of
disorder only states with a stability above some threshold occur.
A measure of the stability is the energy it costs to create a quasielectron-quasihole pair. In the thin limit, this 
decreases monotonously when $q$ increases\cite{bk3}---in good agreement with the hierarchy prediction\cite{halperin84}.
Fig. \ref{figure3} shows $1/q$ for all hierarchy states
in the lowest Landau level. There is a unique state at each rational filling factor, so this is simply a plot of $1/q$ for all 
rational numbers $\nu=p/q$ in the interval $[0,1]$ (where $q$ is odd and $p,q$ are relatively prime).
This is a self-similar, fractal, structure---enlarging any interval of $\nu$ gives an identical figure.   
Since the solvable limit predicts that the gap increases with $1/q$, 
states above some roughly horizontal line should be observed in a given sample. 
In the figure, the states observed in the interval $2/7\le \nu \le 2/3$ 
for an extreme high mobility sample\cite{pan} are indicated.  
The agreement with the prediction is surprisingly good. The deviations seen in Fig. \ref{figure3} may be due to 
corrections to exact particle-hole symmetry and to the difficulty to observe a weak state that is close 
to a much more stable state---this for instance explains why the Jain state at $\nu = 10/21$ is observed, 
but not (yet) the state at $\nu = 7/19$.
The latter, which we predict to be on the verge of observation, is interesting in that it would be the first  
daughter of a non-Jain state. We conclude that Fig. \ref{figure3} in general determines what states 
should be observed at given disorder.

We now discuss the phase diagram in the filling factor-disorder plane in the thin limit,  making two assumptions:
1) The states with a gap above a certain cutoff  are formed at given disorder and this cutoff
decreases monotonously with decreasing amount of disorder, and 2) phase transitions 
between QH-states are caused exclusively by condensation of the quasiparticles discussed 
above. 1) means that  Fig. \ref{figure3} can be thought of as a diagram in the filling factor-disorder plane and 2)
implies that the phase boundary for $\nu=p/q$ must contain precisely those states that can be obtained 
by (successive) condensation of such quasiparticles. For example, the 1/3-dome must 
extend over $\nu= p/(2p+1)\rightarrow 1/2$ and over $\nu= p/(4p-1)\rightarrow 1/4$ but include no larger or smaller $\nu$. 
This gives the phase diagram in Fig. \ref{figure4}, where only the topology of the phases and their relative hights
are significant. The topology is identical to the one in the lowest Landau level part of the global phase diagram of   Refs. \onlinecite{global} and \onlinecite{lutken}.

Given the assumptions, this establishes the global phase diagram as $L\rightarrow 0$. As $L\rightarrow \infty$, 
the TT-state at any $\nu=p/q$, $q$ odd, develops into a QH-state, as will be argued  below. If there is no phase transition when 
$L$ grows, then this state is the ground state in the experimental regime and the QH effect is observed 
at this filling fraction. However, phase transitions may occur and other states 
may be observed.   
We believe that this explains the observed deviations from the global phase diagram, such 
as the insulating phases near Laughlin fractions\cite{jiang}.
In fact, the even denominator fractions are an example of this phase transition scenario. Even though we have excluded 
them from the discussion above, 
the gapped TT-states are the ground states also for these fractions as $L\rightarrow 0$. 
We believe that the well-understood $\nu=1/2$ case is representative: 
As $L$ increases, the gapped TT-state, which has unit cell 10, gives way to a state with gapless 
neutral excitations for $L \sim 5 \ell$---a Luttinger liquid; 
this state then develops smoothly into the observed gapless two-dimensional bulk state\cite{bk1,bk2}. 
This shows that the small $L$ limit may describe also non-hierarchy states.

It is  crucial that our results can be extended from the thin cylinder, were they are established, to the experimentally 
relevant two-dimensional bulk case. First we note that 
the qualitative features of the TT-states and the QH hierarchy states are the same: they have a gap, the 
same quantum numbers and, in particular, they have quasiparticles with the same fractional charge.
Also note, that 
while any QH state on a torus shows a periodic variation in the density\cite{haldanewfs}, the approach to homogeneity is very rapid 
but does not correspond to a phase transition.
The two schemes--the thin limit and the CFT approach--give an identical hierarchy of states, this clearly suggests an adiabatic connection.

For a short-range
interaction the Laughlin states are the ground states for all $L$\cite{hr}; this establishes that the TT-states at these fractions
develop continuously into the bulk QH-states as $L\rightarrow \infty$ without a phase transition. 
Noting that the 
problem on a cylinder with circumference $L$ can equivalently be thought of as the infinite  two-dimensional 
case with an $L$-dependent hamiltonian, we conclude that the TT-state and the bulk QH-state are
adiabatically connected for the Laughlin fractions (and a short range interaction).  
We believe this holds generically for fractions where a QH hierarchy state is observed, {\it ie} this state is adiabatically connected to
the corresponding TT-state. 
This claim is supported by numerical simulations on small systems, where the gap to the first excited state 
has been shown to remain finite for all odd (but for no even) $q\le 11$\cite{bk2}. Moreover, Jain's wave functions for $\nu=n/(2kn\pm 1)$ 
reduce to the appropriate TT-states as $L \rightarrow 0$\cite{bk2}, thus giving 
a strong argument for the adiabatic continuity also for these fractions. However, the status of the fractions that can not 
be interpreted as an integer effect of composite fermions has until now been less clear. 

Around 1990 it was noted that Laughlin's wave functions take the form of  correlation 
functions in certain conformal field theories\cite{fubini, mooreread}, and  
it was also conjectured that this is true for general QH-states\cite{wen,mooreread}.
Recently it was shown that the composite fermion wave functions in the Jain sequence $\nu = n/(2kn+1)$, $k,n=1,2\dots$ can be 
constructed from correlators in a CFT with $n$ bosonic fields\cite{hans}. A natural extension of this 
construction gives, for $n$ bosonic fields, a set of wave functions labeled  by $n$ positive integers $k_i$, $i=1,2,\dots n$. 
If $k_i=1$ for $i=2,\dots n$, then $\nu_n=n/(2k_1n+1)$ and the wave functions are  Jain's composite 
fermion wave functions. 
For a general set $\{k_i\}$, the wave function approaches the TT-ground state in the thin cylinder limit.
These fractions, in the interval $[\frac 1 4, \frac 1 2]$, are indicated   in Fig. \ref{figure4}.
The $k_i$'s determine the densities of the $n$ condensates of quasielectrons that build 
up the state---the composite fermion state is the one where all but the first of these condensates have maximal density.
Using the methods of Ref. \onlinecite{hans} one can construct wave functions also for the pertinent quasihole and quasielectron excitations, and show that the expected
charge and statistics properties of these particles are reflected in the algebraic properties ($U(1)$ charge and commutation relations) of the corresponding anyonic operators.  Details will be published later, and here we only present the $n=2$ 
wave functions for $N=2k_2 M_2$ particles at positions $z_i = x_i + iy_i$: 

\beqa \label{wavefns}
\Psi \!=\!\! \sum_{i_1<i_2 \dots < i_{M_2}}  (-1)^{\sum_m i_m} \partial_{i_1}\!\cdots \!\partial_{i_{M_2}} \!\left[   \prod_{i_l,\bar i_m}(z_{i_l}\!-\!z_{\bar i_m})^{2k_1} \right.  \nonumber
\\    \times \left.
\prod_{i_l<i_{m}}(z_{i_l}\!-\!z_{i_m})^{2(k_1+k_2)-1}\! \prod_{\bar i_l<\bar i_m}(z_{\bar i_l}\!-\!z_{\bar i_m})^{2k_1+1} \right]\! e^{-\sum_i \frac { |z_i|^2} {4\ell^2}} \  , \nonumber
\eeqa
where  $\{i_m\}$ is a subset of $M_2$ indices, and $\{\bar i_m\}$ are the 
remaining $M_1=(2k_2-1)M_2$ indices.  For $k_2=1$, $\Psi$ are Jain's wave functions at $\nu = 2/(4k_1+1)$, while $k_1=1,k_2=2$ gives our proposal for 
the observed state at $\nu = 4/11$. The thin limits\cite{bk2} of the $n=2$ states are the TT-states $(10_{2k_1})_{2k_2-1}10_{2k_1-1}$.

We compared our candidate wave function at $\nu=4/11$ with exact 
diagonalization results in the disc geometry. For eight particles, the overlap with the exact ground state 
is 0.72 and the energy $E_{4/11} = 5.317$ is in good agreement with the exact one, 
$E_{exact}= 5.297 $ (in units of $e^2/\epsilon \ell$ ). 
Preliminary results for  a torus, where there are no edge effects,  are encouraging.

In summary, we presented theoretical support for the global phase diagram and obtained an explicit 
realization of the hierarchy of fractional QH-states in a solvable limit.  For a large class
of states, including those of Laughlin and Jain,
we presented explicit many body wave functions that represent
an adiabatic continuation from the solvable limit to the experimentally
relevant region.

We thank  Jainendra Jain, Steven Kivelson, Juha Suorsa and  Susanne Viefers for  helpful discussions and Chia-Chen Chang for help with the numerical calculations.    
This work was supported by the Swedish Research Council and by NordForsk.


\begin{thebibliography}{99}

{\footnotesize


\bibitem{laughlin81} R. B.  Laughlin, Phys. Rev. B. {\bf 23}, 5632, (1981); 
B. I.  Halperin,  Phys. Rev. B {\bf 25}, 2185, (1982).

\bibitem{laughlin83} R. B.  Laughlin,  Phys. Rev. Lett. {\bf 50}, 1395 (1983).

\bibitem{haldane83}F. D. M.  Haldane, Phys. Rev. Lett. {\bf 51}, 605 (1983).

\bibitem{halperin84} B. I.  Halperin,  Phys. Rev. Lett. {\bf 52}, 1583, 2390(E) (1984). 

\bibitem{jain89}  J. K. Jain, Phys. Rev. Lett. {\bf 63}, 199 (1989).

\bibitem{jainbook} J. K. Jain,  {\it Composite fermions}, (Cambridge University Press, 2007). 

\bibitem{read90} N. Read, Phys. Rev. Lett. {\bf 65}, 1502 (1990);
B. Blok, and  X. G. Wen, Phys. Rev. B {\bf 43}, 8337 (1991).

\bibitem{greiter} M. Greiter, Phys. Lett. {\bf B 336}, 48 (1994). 

\bibitem{global} S. A. Kivelson, D.-H. Lee, and S.-C. Zhang,  Phys. Rev. B {\bf 46}, 2223 (1992). 

\bibitem{lutken} C. A.  L\"utken, and  G. G. Ross, Phys. Rev. B {\bf 45}, 11837 (1992); 
Phys. Rev. B {\bf 48}, 2500 (1993).

\bibitem{jiang} H. W. Jiang, {\it et al}.,  Phys. Rev. Lett. {\bf 65}, 633 (1990);
H. C. Manoharan, and M.  Shayegan,  Phys. Rev. B {\bf 50}, 17662 (1994).

\bibitem{pan} W. Pan,  {\it et al}.,  Phys. Rev. Lett. {\bf 90}, 016801 (2003).

\bibitem{mani} R. G. Mani, and K.  Z.  von Klitzing, Phys. B {\bf 100}, 635 (1996);
M. O. Goerbig, P. Lederer, and C. Morais Smith,  Europhysics Letters {\bf 68}, 72 (2004).

\bibitem{other} J. K. Jain and V. J. Goldman,  Phys. Rev. B {\bf 45}, 1255 (1992); 
J. N. Ginocchio and W. C. Haxton, Phys. Rev. Lett. {\bf 77}, 1568 (1996);
P. Sitko, K.-S. Yi and J. J. Quinn, Phys. Rev. B {\bf 56}, 12417 (1997).

\bibitem{bk2} E. J. Bergholtz, and A. Karlhede,   J. Stat. Mech. (2006) L04001.

\bibitem{tt} R. Tao, and D. J. Thouless,  Phys. Rev. B {\bf 28}, 1142 (1983).

\bibitem{su} W. P. Su,  Phys. Rev. B {\bf 30}, 1069 (1984).

\bibitem{bk3} E. J. Bergholtz, and A. Karlhede,   arXiv:0712.1927 (2007). 

\bibitem{bk1} E. J. Bergholtz, and A. Karlhede,  Phys. Rev. Lett. {\bf 94}, 26802 (2005).

\bibitem{haldanewfs}  F. D. M. Haldane, and E. H. Rezayi, Phys. Rev. B {\bf 31},
2529 (1985).

\bibitem{hr} E. H.  Rezayi, and  F. D. M.  Haldane, Phys. Rev. B {\bf 50}, 17199 (1994).

\bibitem{mooreread} G. Moore,  and N.  Read,   Nucl. Phys. B {\bf 360}, 362 (1991).

\bibitem{fubini} S. Fubini,  Mod. Phys. Lett. A 6, 347 (1991). 

\bibitem{wen} X.-G. Wen,  Advances in Physics {\bf 44}, 405 (1995).

\bibitem{hans} T. H. Hansson, C.   C. Chang, J. K.  Jain,  and S. F. Viefers,  Phys. Rev. Lett. {\bf 98} 076801 (2007), arXiv:0704.0570  (2007).


}

\end{thebibliography}
\end{document}